\documentclass[twocolumn, trackchanges]{aastex6}

\usepackage{multirow}
\newcommand{\rl}{\raggedleft}
\usepackage{url}

\usepackage{xcolor}
\usepackage{amsmath}

\begin{document}


\title[Parameter estimation for GW bursts with BayesWave]{Parameter estimation for gravitational-wave bursts with the BayesWave pipeline}


\author{Bence B\'ecsy\altaffilmark{1,2}}
\author{Peter Raffai\altaffilmark{1,2}}
\author{Neil J. Cornish\altaffilmark{3}}
\author{Reed Essick\altaffilmark{4}}
\author{Jonah Kanner\altaffilmark{5}}
\author{Erik Katsavounidis\altaffilmark{4}}
\author{Tyson B. Littenberg\altaffilmark{6}}
\author{Margaret Millhouse\altaffilmark{3}}
\and
\author{Salvatore Vitale\altaffilmark{4}}


\altaffiltext{1}{Institute of Physics, E\"otv\"os University, 1117 Budapest, Hungary; \href{mailto:becsybence@caesar.elte.hu}{becsybence@caesar.elte.hu}}
\altaffiltext{2}{MTA-ELTE EIRSA ``Lend\"ulet'' Astrophysics Research Group, 1117 Budapest, Hungary}
\altaffiltext{3}{Department of Physics, Montana State University, Bozeman, MT 59717, USA}
\altaffiltext{4}{Massachusetts Institute of Technology, 185 Albany St, 02139 Cambridge USA}
\altaffiltext{5}{LIGO Laboratory, California Institute of Technology, Pasadena, CA 91125, USA}
\altaffiltext{6}{NASA Marshall Space Flight Center, Huntsville AL  35812, USA}

\begin{abstract}
We provide a comprehensive multi-aspect study on the performance of a pipeline used by the LIGO-Virgo Collaboration for estimating parameters of gravitational-wave bursts. We add simulated signals with four different morphologies (sine-Gaussians, Gaussians, white-noise bursts, and binary black hole signals) to simulated noise samples representing noise of the two Advanced LIGO detectors during their first observing run. We recover them with the BayesWave (BW) pipeline to study its accuracy in sky localization, waveform reconstruction, and estimation of model-independent waveform parameters. BW localizes sources with a level of accuracy comparable for all four morphologies, with the median separation of actual and estimated sky locations ranging from 25.1$^{\circ}$ to 30.3$^{\circ}$. This is a reasonable accuracy in the two-detector case, and is comparable to accuracies of other localization methods studied previously. As BW reconstructs generic transient signals with sine-Gaussian wavelets, it is unsurprising that BW performs the best in reconstructing sine-Gaussian and Gaussian waveforms. BW's accuracy in waveform reconstruction increases steeply with network signal-to-noise ratio (SNR$_{\rm net}$), reaching a $85\%$ and $95\%$ match between the reconstructed and actual waveform below SNR$_{\rm net} \approx 20$ and SNR$_{\rm net} \approx 50$, respectively, for all morphologies. BW's accuracy in estimating central moments of waveforms is only limited by statistical errors in the frequency domain, and is affected by systematic errors too in the time domain as BW cannot reconstruct low-amplitude parts of signals overwhelmed by noise. The figures of merit we introduce can be used in future characterizations of parameter estimation pipelines.
\end{abstract}




\section{Introduction} \label{sec:intro}

The network of Advanced LIGO (aLIGO) gravitational-wave (GW) detectors \citep{aligo}, consisting of aLIGO-Hanford (H1) and aLIGO-Livingston (L1), finished its first observing run (O1) in January 2016. During O1, this network achieved the first direct detections of GWs by detecting GW150914  \citep{detection} and GW151226 \citep{GW151226}, two signals from coalescences of binary black holes. Besides binary black holes, other astrophysical sources of GW transients (e.g.~core-collapse supernovae, magnetar flares, and cosmic string cusps) are also targeted by aLIGO \citep{prospects}. Searches for generic GW transients aim to detect weakly-modeled GW signals (``bursts'') from such systems, as well as from binary black holes, and also from as-yet-unknown sources (see e.g.~\citealt{burst_comp}).

Detections of GW signals will be used to test and constrain models of astrophysical sources (see e.g.~\citealt{astro_comp, belczynski}). This usually requires reconstructing the signal waveform from the GW detector output and estimating parameters of the waveform (see e.g.~\citealt{pe_comp}). For sources where an accurate waveform model exists, such as binary black holes in circular orbits, this is done by matching the detector output with template waveforms (see e.g.~\citealt{pe_comp}). In this case, the estimated parameters are astrophysical, e.g., chirp mass and spins. Parameter estimation (PE) for burst signals, where no model templates exist, need a different approach. In such cases, basis functions are used to reconstruct the waveform and to estimate model-independent parameters of it, such as central time and frequency, signal duration and bandwidth. Besides these intrinsic parameters of the waveform, estimates can also be given on the extrinsic parameters of the source (e.g.~sky location).

BayesWave (BW) is a pipeline for detecting and characterizing GW bursts, that works within the framework of Bayesian statistics and uses sine-Gaussian wavelets as basis functions to reconstruct the signal \citep{BW}. In O1, BW was used as a follow-up PE tool on triggers provided by the coherent Waveburst (cWB) search pipeline \citep{cWB,cWB2}, which identifies coincident excess power in strain data of multiple GW detectors. Note however, that cWB can also reconstruct the sky location of a GW source and the waveform of the GW signal, independently from BW \citep{cWB_PE}. This provides an opportunity to compare the performances of BW and cWB in PE using the same set of triggers (for the results of this comparison, see Section \ref{ssec:skyloc}). BW is effective in distinguishing GW signals from non-Gaussian noise artifacts (``glitches''), which enables the combination of the cWB and BW pipelines to achieve high-confidence detections across a range of waveform morphologies \citep{BW_det,BW_wfcompl}. The estimates of mass parameters and sky location obtained by BW for GW150914 have shown to be consistent with template-based PE pipelines \citep{burst_comp}.

In this paper we characterize BW's performance in PE by injecting a large set of simulated signals into simulated aLIGO noise, and recovering them and their parameters with BW. The main purpose of this study is to determine the accuracy of these reconstructions that can be achieved with BW. By knowing the accuracy, future studies can identify the broadest range of astrophysical models that can be tested with BW, while further improvements of BW can be guided by these results. Among the estimated parameters, we give special attention to sky location of the GW source, because of its key role in electromagnetic (EM) follow-up observations of GW events (see e.g.~\citealt{EMfup_comp, singer_etal, BNS_PE, CBCBurstSkyloc}). Sky localization of GW burst sources can be carried out with the cWB and LALInferenceBurst (LIB) pipelines (\citealt{oLIB,LALInference}) too. An extensive analysis on the sky localization performance of cWB and LIB was published in \citet{skyloc}. Here we present a similar analysis for BW in order to characterize its performance and to allow comparisons with other burst pipelines studied in \citet{skyloc}. Note however, that as we use a reduced set of triggers compared to \citet{skyloc} (for an explanation, see Appendix \ref{sec:app}), our results in Figures 1-4 should not be compared directly with results in Figures 3-6 of \citet{skyloc}. Instead, to allow direct comparisons between BW, cWB, and LIB, we repeat our analysis with cWB and with LIB on the same reduced set of triggers, and present the results in Figure Sets 1-4 (available in the online journal). Also note that new cWB sky localization results for binary black holes presented recently (see \citealt{CBCBurstSkyloc}) show that cWB's performance has improved significantly for a three-detector network, while it has not changed significantly for the two-detector case we present here.

We focus on three aspects of BW's performance: (i) sky localization, (ii) waveform reconstruction, and (iii) estimation of model-independent waveform parameters. In Section \ref{sec:methods}, we describe the methods used for creating simulated signals and noise samples, and used by BW for carrying out PE. In Section \ref{sec:results}, we present results of our analyses regarding all (i)-(iii) aspects. We summarize our findings and highlight some implications in Section \ref{sec:concl}.

\section{Methods} \label{sec:methods}

We used software injections to test the PE performance of BW, i.e.~we created mock samples of aLIGO noise and added simulated GW signals with four different morphologies to these samples. We then used these samples at trigger times provided by cWB as inputs for BW to test what it recovers from the signals embedded in the mock detector noise. In this section we discuss the characteristics of noise samples and of simulated signals we used (Section \ref{ssec:data}), as well as methods BW uses for PE (Section \ref{ssec:bw}).

\subsection{Noise and injections} \label{ssec:data}

In this section we summarize characteristics of injections and noise samples we used in our analyses, which are the same as the ones used in \citet{skyloc}. For further details on this see Section 2, Appendix C and Table 4 in \citet{skyloc}.

In our analysis we considered a two-detector network consisting of H1 and L1. We used stationary, Gaussian mock noise samples generated using the expected 2015 sensitivity curve of aLIGO, thus they have slightly different characteristics than the actual noise collected during the O1 run. Projections show that the two LIGO detectors will operate in the first two months of the second observing run (O2) with similar sensitivity curves they operated with during O1. Thus, we expect that our results are representative for this first period of O2 as well.

Our set of software injections consists of signals with four different morphologies: sine-Gaussians (\textit{SG}); Gaussians (\textit{G}); white-noise bursts (\textit{WNB}); and binary black hole (\textit{BBH}) mergers. This wide range of signal morphologies allows us to test the PE performance of BW with minimal assumptions on the GW signal. The amplitude distribution of injected signals was chosen such as to represent a uniform distribution of GW sources in volume. Signal injections were distributed uniformly over the sky and were regularly spaced in time.

The number of signals we analyzed was determined by multiple factors (see Table \ref{tab:injections}): (i) the BW version we used runs only on triggers produced by cWB \citep{burst_comp}; (ii) we reduced the number of \textit{BBH} triggers in order to reduce computational costs; (iii) we only used signals correctly identified  as signals by BW. For details on why BW identified many \textit{SG} and \textit{WNB} signals as glitches or Gaussian noise, and how this has been improved for O2, see Appendix \ref{sec:app}.

\floattable
\begin{deluxetable*}{c|ccccc}
\tablecaption{Number of injected signals for each morphology at different stages of the analysis. For details on why BW identified many \textit{SG} and \textit{WNB} signals as glitches or Gaussian noise, and how this has been improved for O2, see Appendix \ref{sec:app}. \label{tab:injections}}
\tablecolumns{4}
\tablewidth{0pt}
\tablehead{
\colhead{ } & \colhead{\textit{SG}} & \colhead{\textit{G}} & \colhead{\textit{WNB}} & \colhead{\textit{BBH}} 
}
\startdata
Triggers produced by cWB & 1112 & 256 & 769 & 2488\\
Left out to reduce computational costs & 0 & 0 & 0 & -1988\\
\hline
Analyzed by BW & 1112 & 256 & 769 & 500\\
Identified as glitches or Gaussian noise by BW & -779 & 0 & -355 & -1\\
\hline
\textbf{Used in our analysis} & \textbf{333} & \textbf{256} & \textbf{414} & \textbf{499}\\
\enddata
\end{deluxetable*}

\textit{SG} waveforms are often used to model generic transients (e.g.~\citealt{SG}), because they are the most localized signals in time-frequency space where generic burst searches (including cWB) operate (see \citealt{shourov}). We define \textit{SG} waveforms with the following two equations:

\begin{subequations}
\begin{eqnarray}
h_{+} (t) = \cos (\alpha) h_{\rm rss} \sqrt{\frac{4 f_0 \sqrt{\pi}}{Q(1+\cos (2\phi_0)e^{-Q^2})}} \nonumber\\ \times \cos (2 \pi f_0 (t - t_0) + \phi_0) e^{-(t-t_0)^2/\tau^2}
\label{eq:SG+}
\end{eqnarray}
\begin{eqnarray}
h_{\times} (t) = \sin (\alpha) h_{\rm rss} \sqrt{\frac{4 f_0 \sqrt{\pi}}{Q(1-\cos (2\phi_0)e^{-Q^2})}} \nonumber\\ \times \sin (2 \pi f_0 (t - t_0) + \phi_0) e^{-(t-t_0)^2/\tau^2},
\label{eq:SGx}
\end{eqnarray}
\end{subequations}
where $\alpha \in [0, \pi/2]$ is a parameter that sets the relative weights between polarizations $h_+$ and $h_{\times}$, ${h_{\rm rss}^2=\int (h_+^2 + h_{\times}^2) \rm{d} t}$ is the square of the root-sum-squared strain amplitude chosen as a free parameter in the amplitude randomization process, $f_0$ is the central frequency, $t_0$ is the central time, $\phi_0$ is the phase at time $t=t_0$, $\tau$ is the width of the signal in the time domain, and $Q = \sqrt{2} \pi \tau f_0$ is the quality factor encoding the characteristic number of cycles within duration of the signal.

\textit{G} signals are the special cases of \textit{SG} signals when $f_0 \to 0$, and are defined as:

\begin{subequations}
\begin{equation}
h_{+} (t) = \cos (\alpha) \frac{h_{\rm rss}}{\sqrt{\tau}} \left( \frac{2}{\pi} \right)^{1/4} e^{-(t-t_0)^2/\tau^2}
\label{eq:G+}
\end{equation}
\begin{equation}
h_{\times} (t) = \sin (\alpha) \frac{h_{\rm rss}}{\sqrt{\tau}} \left( \frac{2}{\pi} \right)^{1/4} e^{-(t-t_0)^2/\tau^2}.
\label{eq:Gx}
\end{equation}
\end{subequations}
Despite their similarity to \textit{SG}s, these signals pose different challenges, because they have their highest amplitude at $f=0$ Hz in frequency domain, and thus they have most of their power at low frequencies where aLIGO is less sensitive.
 
\textit{WNB} waveforms are intended to model a time-localized excess power uniformly distributed in a given frequency band, and satisfy:

\begin{equation}
h_{+,\times}(t) \propto e^{\frac{-(t-t_0)^2}{\tau^2}} \int_{-\infty}^{\infty} e^{-i 2\pi f t} w(f) {\rm d} f,
\label{eq:WNB}
\end{equation} 
where $w(f)$ values are randomly drawn from a Gaussian white noise within and chosen to be $w(f) = 0$ outside the band $f\in[f_{\rm min},f_{\rm max}]$. We generated the right side of equation (\ref{eq:WNB}) independently for the + and $\times$ polarizations, and normalized them to get $h_+$ and $h_{\times}$ with the desired $h_{\rm rss}$. Unlike signals with the other three morphologies, \textit{WNB} signals are not elliptically polarized, because the procedure used to produce them generates $h_+$ and $h_{\times}$ independently.
 
The only astrophysical signals we used were \textit{BBH}s with spins aligned or anti-aligned with the orbital angular momentum. We only considered binaries with relatively high detector-frame total masses ($M_{\rm tot} \in [30, 50] \ M_{\odot}$), because their signals are more compact in time-frequency space, which makes them good targets for generic burst searches. Three different methods have been used for calculating the waveform in the three different phases of binary evolution: 3.5PN post-Newtonian expansion, numerical relativity, and analytic quasi-normal modes to calculate the inspiral, merger, and ringdown waveforms, respectively (see \citealt{imr_waveforms1} and \citealt{imr_waveforms2} for details).

\subsection{The BayesWave pipeline} \label{ssec:bw}

BW uses a trans-dimensional Reversible Jump Markov Chain Monte Carlo (RJMCMC) algorithm \citep{rjmcmc} to explore the following three competing models of the data, and test them with the input data samples from each aLIGO detector: i) Gaussian noise only; ii) Gaussian noise with glitches; iii) Gaussian noise with a GW signal. This approach makes BW effective in distinguishing GW signals from glitches \citep{BW_det}, but it also makes BW computationally expensive, and thus in O1 BW was used to follow-up candidate events from cWB.

BW assumes that all signals are elliptically polarized, i.e.~$h_{\times}=\epsilon h_{+} e^{i\pi/2}$, where $\epsilon \in [0,1]$ is the ellipticity parameter, which is 0 for linearly polarized signals and 1 for circularly polarized ones. This is a valid assumption for many expected astrophysical signals, but not for our injections with \textit{WNB} morphology (see Section \ref{ssec:data}). However, for a LIGO-only network, it is often the case that only a single combination of the two polarizations, rather than the separate $+$ and $\times$ components, will be detectable, making the elliptical constraint a fair approximation for many cases.

We used the BW version which had been used for the offline analysis of O1 data to attain a characterization of BW's performance during O1, and to support a fair comparison with the versions of other PE pipelines characterized in \citet{skyloc}. PE pipelines used by the LIGO-Virgo Collaboration (including BW) have undergone improvements since the beginning of O1 (some of which were motivated by this study).

\section{Results} \label{sec:results}

In this section we show how BW performed in different aspects of PE. These aspects are sky localization (see Section \ref{ssec:skyloc}), waveform reconstruction (see Section \ref{ssec:wfrec}), and point estimates of waveform central moments (see Section \ref{ssec:pointest}).

Even though BW's current (O2) version is more efficient in identifying signals (see Appendix \ref{sec:app}), we used the version of BW used during O1 in order to characterize BW's performance during O1, and to allow a comparison of our results with the ones presented in \citet{skyloc}. We only analyzed signals that were properly identified as signals by BW (see Table \ref{tab:injections}). We present a reproduction of results of \citet{skyloc} for the subset of events we used in this study, to enable a fair comparison of sky localization results (see Figure Sets 1-4.).

Results presented here depend on the parameter distributions of injected signals defined in Table 4 of \citet{skyloc}, and on the corresponding detection efficiencies of the combination of cWB and BW pipelines for the different parameter sets. Results are particularly dependent on the chosen $h_{\rm rss}$ distribution of injected signals, and thus on the network signal-to-noise ratio (SNR$_{\rm net}$) distribution of them (see inset of Figure \ref{fig:overlap_hist}). However, the $h_{\rm rss}$ distribution we chose for this study is a good approximation for generic burst signals uniformly distributed in volume (see Appendix C in \citealt{skyloc}).

\subsection{Sky localization} \label{ssec:skyloc}

BW computes a skymap defined as the posterior probability density function of the GW source location expressed as a function of celestial coordinates $\alpha$ (right ascension) and $\delta$ (declination), denoted by $p_{\rm sky} (\alpha, \delta)$. Example skymaps for each morphology are shown in Appendix \ref{sec:exsky}. Skymaps for all the injections can be found in the Burst First2Years sky localization Open Data release\footnote{\url{http://www.ligo.org/scientists/burst-first2years/}}. There are many possible quantitative measures for the ``goodness'' of source localization, here we implement the ones defined in \citet{skyloc}, i.e.~\emph{angular offset}, \emph{searched area}, \emph{extent} and \emph{fragmentation}. We reproduced results of \citet{skyloc} for LIB and cWB using the same subset of events we used in this study (the ones identified as signals by BW) to enable a direct comparison of the results (see Figure Sets 1-4).

\begin{figure}
\figurenum{1}
\epsscale{1.2}
\plotone{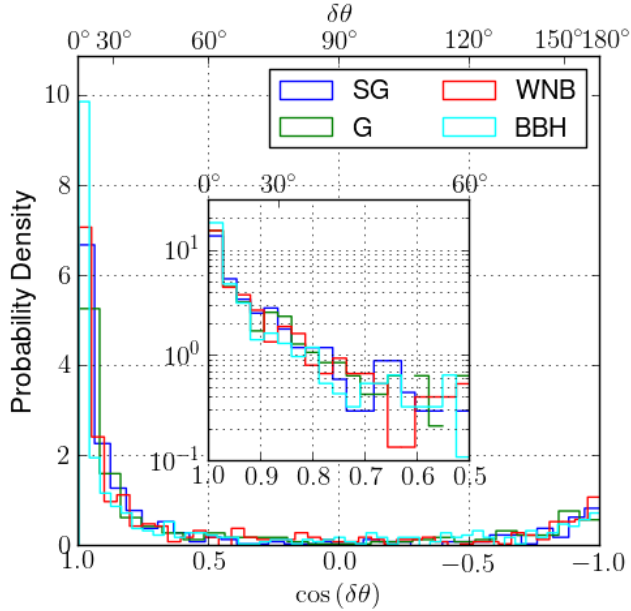}
\caption{Normalized histograms of \emph{angular offsets} ($\delta \theta$) for injections with four different morphologies (\textit{SG}, \textit{G}, \textit{WNB}, \textit{BBH}). Most of the injected signals have $\cos(\delta \theta) = 1$, which indicates that BW tends to place the most probable location close to the true location. Note that the distributions for different morphologies are very similar to each other, which means that the angular offset does not depend strongly on signal morphology. The complete figure set (3 figures) showing the same plot for cWB and LIB pipelines is available in the online journal. \label{fig:dtheta}}
\end{figure}

The first measure is the \emph{angular offset} ($\delta \theta$), which is the angular distance between the maximum of $p_{\rm sky}$ and the true location of the injected signal. Figure \ref{fig:dtheta} shows normalized histograms of $\cos (\delta \theta)$ for all injections, with the upper axis showing the corresponding $\delta \theta$ values. The distribution has a peak at $\cos(\delta \theta) = 1$, which suggests that BW tends to reconstruct the most probable location of the source close to the actual source location. There is also a smaller peak at $\cos(\delta \theta) = -1$, which indicates that it is more likely that BW reconstructs the opposite direction of the sky compared to the location of the injected signal than a direction perpendicular to the injected signal's location. This is due to the fact that opposite directions cannot be distinguished using the network antenna pattern which has the same value at opposite directions because of the near co-alignment of H1 and L1 detectors \citep{singer_etal}. However, the peak at $\cos(\delta \theta) = -1$ is smaller than the one at $\cos(\delta \theta) = 1$ because opposite directions are only allowed by the triangulation ring when the source is right above (or below) the detectors, and thus the triangulation ring is a great circle on the celestial sphere. Note that the distributions for different morphologies are very similar to each other, which means that the angular offset depends weakly on signal morphology. We show summary statistics of $\delta\theta$ distributions for all morphologies in Table \ref{table:skymap_stat}. It is clearly visible that BW performs best for \textit{BBH} signals, while \textit{SG}, \textit{G} and \textit{WNB} signals show slightly larger $\delta\theta$ values. Statistical errors on reported values are in the order of a few percent. Figure Set 1 shows normalized histograms of $\cos (\delta \theta)$ obtained with the cWB and LIB pipelines on the subset of signals identified as signals by BW.

EM follow-up observations tend to target the point of the sky with the highest $p_{\rm sky}$ value first, and continue with points having lower $p_{\rm sky}$ values. This motivates the introduction of the \emph{searched area} ($\mathcal{A}$) as a second measure, which is the total sky area observed before aiming a hypothetical telescope at the true location of the source:

\begin{equation}
\mathcal{A} = \int H(p_\mathrm{sky} (\alpha, \delta) - p_0) \ \mathrm{d}\Omega
\label{eq:searchedarea}
\end{equation}
where $H$ is the Heaviside step function, $p_0$ is the value of $p_{\rm sky}$ at the true location of the source, and $\rm{d} \Omega = \cos \delta \ \rm{d} \delta \ \rm{d} \alpha$.

\begin{figure}
\figurenum{2}
\epsscale{1.2}
\plotone{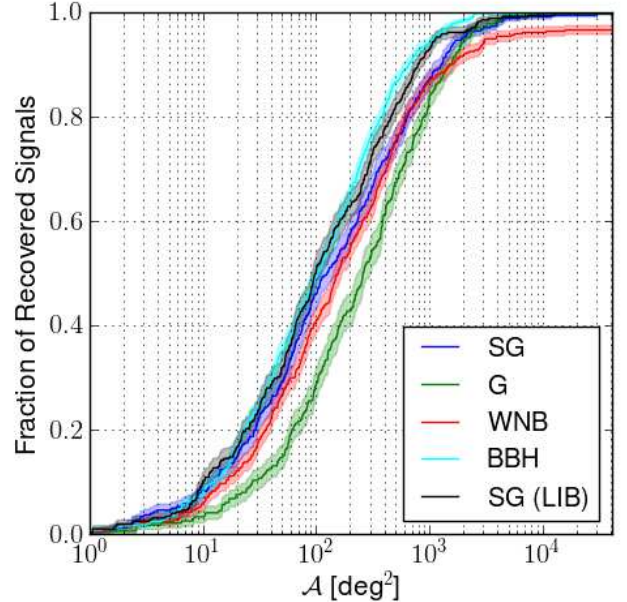}
\caption{Cumulative histograms of \emph{searched area} ($\mathcal{A}$). Histograms for different morphologies follow a similar trend, except that the curves are shifted along the horizontal axis. A reference curve labeled with SG (LIB) shows results for the LIB pipeline on the subset of \textit{SG} signals identified as signals by BW. The complete figure set (3 figures) showing the same plot for cWB and LIB pipelines is available in the online journal. \label{fig:searched_area}}
\end{figure}

We show the cumulative histogram of $\mathcal{A}$ for all injections in Figure \ref{fig:searched_area}. Histograms for different morphologies follow a similar trend, but the curves are shifted along the horizontal axis. This can be quantified e.g.~with median searched area, which is 252.8 deg$^2$ for \textit{G}, 151.0 deg$^2$ for \textit{WNB}, 121.3 deg$^2$ for \textit{SG}, and 99.2 deg$^2$ for \textit{BBH} signals. Another difference between morphologies is that there is a fraction of \textit{WNB} signals with searched area equal to the whole sky ($\mathcal{A} \simeq 4 \cdot 10^4 \ \mathrm{deg}^2$). This is due to the fact that $p_0 = 0$ for these signals, i.e.~the posterior distribution has no support at the true location of the source. There are no such signals with \textit{SG}, \textit{G} and \textit{BBH} morphologies. A reference curve labeled with SG (LIB) shows results for the LIB pipeline on the subset of \textit{SG} signals identified as signals by BW. Note that LIB uses a single sine-Gaussian to reconstruct the signal, so for \textit{SG} injections LIB becomes a matched-filtering analysis for which better performance is expected, while BW sometimes uses more than one sine-Gaussian, because it favors more complex signals. It shows that LIB performed similarly, but slightly better for \textit{SG} signals. We show summary statistics of $\mathcal{A}$ distributions for all morphologies in Table \ref{table:skymap_stat}. It is clearly visible that BW performs best for \textit{BBH} signals, while \textit{SG}, \textit{G} and \textit{WNB} signals show significantly larger $\mathcal{A}$ values. Statistical errors on reported values are in the order of a few percent. Figure Set 2 shows normalized histograms of $\mathcal{A}$ obtained with the cWB and LIB pipelines on the subset of signals identified as signals by BW.

\begin{figure}
\figurenum{3}
\epsscale{1.2}
\plotone{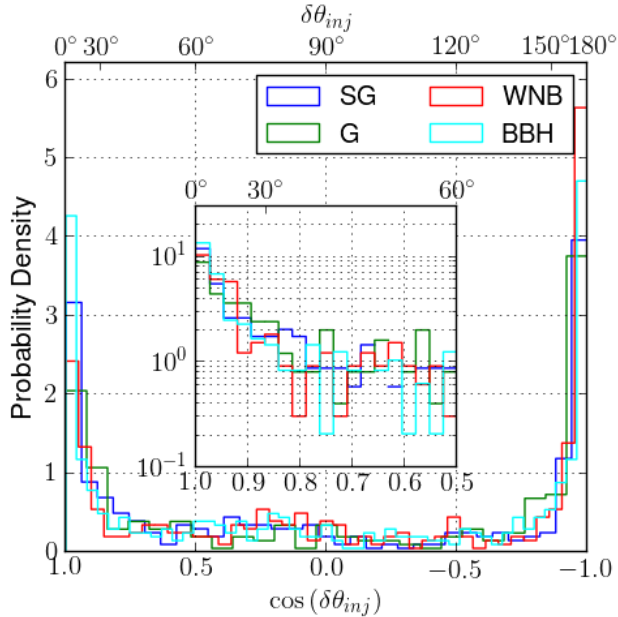}
\caption{Normailzed histograms of the \emph{extent} ($\delta \theta_{\rm inj}$) of skymaps for the four different injection morphologies. The shown distributions are bimodal for all morphologies with peaks at $\cos (\delta \theta_{\rm inj}) = \pm 1$. The complete figure set (3 figures) showing the same plot for cWB and LIB pipelines is available in the online journal. \label{fig:extent}}
\end{figure}

Even if $\delta \theta$ and $\mathcal{A}$ are small, the favored sky positions can still be either well localized or spread out over various parts of the sky. To quantify this feature, we introduce the \emph{extent} ($\delta \theta_{\rm inj}$) of a skymap as the maximum angular distance between the location of the injected signal and any other point satisfying $p_{\rm sky} (\alpha,\delta) \geq p_0$. We show histograms of $\delta \theta_{\rm inj}$ in Figure \ref{fig:extent}. The shown distributions are clearly bimodal with peaks at $\cos (\delta \theta_{\rm inj}) = \pm 1$. The peak at $\cos (\delta \theta_{\rm inj}) = 1$ corresponds to well localized signals, while the peak at $\cos (\delta \theta_{\rm inj}) = -1$ shows that there is a similarly large fraction of events with the skymap extended even to the opposite direction of the sky compared to the true location of the signal. This is due to the same effect described previously when explaining Figure \ref{fig:dtheta}. Note that there are significant differences in the height of the two peaks, e.g. histogram for \textit{BBH} signals have twice as high peak at $\cos (\delta \theta_{\rm inj}) = 1$ than the histogram for \textit{G} signals. Figure Set 3 shows histograms of $\delta \theta_{\rm inj}$ obtained with the cWB and LIB pipelines on the subset of signals identified as signals by BW.

\begin{figure}
\figurenum{4}
\epsscale{1.2}
\plotone{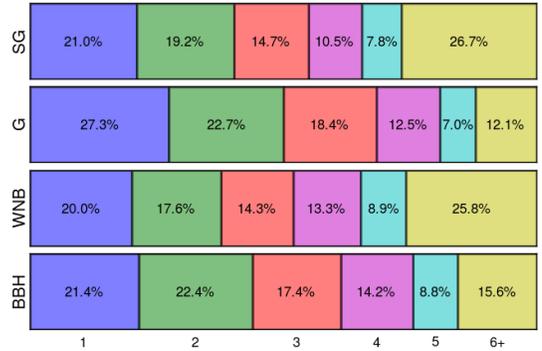}
\caption{Distributions of \emph{fragmentation}. Each row corresponds to one of the four morphologies (\textit{SG}, \textit{G}, \textit{WNB}, \textit{BBH}). Numbers at the bottom of the chart represent the number of disjoint regions in parts of the sky where $p_{\rm sky}\geq p_0$. Number of disjoint regions is less than 4 for more than 50\% of injected signals for all morphologies. The complete figure set (3 figures) showing the same plot for cWB and LIB pipelines is available in the online journal.
\label{fig:fragment}}
\end{figure}

Even if previous measures indicate a well localized source, the skymap can still be fragmented, which makes it more difficult to cover the whole with EM observations. We thus introduce the \emph{fragmentation} of a skymap as the number of disjoint regions in the union of points satisfying $p_{\rm sky}(\delta, \alpha)\geq p_0$. We show the distribution of the number of disjoint regions in Figure \ref{fig:fragment}. Number of disjoint regions is less than 4 for more than 50\% of injected signals for all morphology. Skymaps for \textit{SG} and \textit{WNB} signals are significantly more fragmented than for \textit{G} and \textit{BBH} signals. This is due to the fact that the skymaps of these signals are more likely to have ``fringe peaks''. These are separate rings in the sky corresponding to local maxima of matches between different data streams obtained when they are shifted by half-integer multiples of the period of the signal (for details see Appendix \ref{sec:app}). Figure Set 4 shows distributions of the number of disjoint regions obtained with the cWB and LIB pipelines on the subset of signals identified as signals by BW.

\floattable
\begin{table*}
  \caption{Summary statistics of $\mathcal{A}$ and $\delta\theta$ distributions. Statistical errors are in the order of a few percent.}
  \begin{center}
  \begin{tabular}{p{2.0cm} r||c|c|c|c}
  \hline
  \hline
  \multicolumn{2}{c||}{morphology}                                                  & \textit{BBH} & \textit{SG} & \textit{G} & \textit{WNB} \\
  \hline
  \hline
  \multirow{6}{2cm}{fraction (in \%) with searched area less than}  & \rl5 deg$^2$  & $3.6$ & $4.8$ & $2.3$ & $2.7$ \\
  \multirow{6}{2cm}{}  & \rl20 deg$^2$                                              & $17.4$ & $15.6$ & $7.8$ & $12.3$ \\ 
  \multirow{6}{2cm}{}  & \rl100 deg$^2$                                             & $50.1$ & $46.5$ & $29.3$ & $41.3$ \\
  \multirow{6}{2cm}{}  & \rl200 deg$^2$                                             & $66.5$ & $58.6$ & $43.4$ & $56.0$ \\
  \multirow{6}{2cm}{}  & \rl500 deg$^2$                                             & $87.0$ & $75.4$ & $67.6$ & $76.1$ \\
  \multirow{6}{2cm}{}  & \rl1000 deg$^2$                                            & $94.6$ & $87.7$ & $84.4$ & $87.4$ \\
  \hline
  \multirow{6}{2cm}{fraction (in \%) with $\delta\theta$ less than}  & \rl1$^\circ$ & $3.0$ & $1.2$ & $1.6$ & $1.0$ \\
  \multirow{6}{2cm}{}  & \rl5$^\circ$                                               & $15.4$ & $10.5$ & $12.1$ & $10.1$ \\ 
  \multirow{6}{2cm}{}  & \rl15$^\circ$                                              & $37.5$ & $31.2$ & $30.9$ & $30.2$ \\
  \multirow{6}{2cm}{}  & \rl45$^\circ$                                              & $62.7$ & $69.1$ & $62.9$ & $61.4$ \\
  \multirow{6}{2cm}{}  & \rl60$^\circ$                                              & $69.1$ & $75.7$ & $68.4$ & $67.1$ \\
  \multirow{6}{2cm}{}  & \rl90$^\circ$                                              & $76.4$ & $79.9$ & $75.4$ & $76.1$ \\
  \hline
  \multicolumn{2}{r||}{median searched area}                                        & 99.2 deg$^2$ & 121.3 deg$^2$ & 252.8 deg$^2$ & 151.0 deg$^2$ \\
  \multicolumn{2}{r||}{median $\delta\theta$}                                       & 25.1$^\circ$  & 26.2$^\circ$  & 29.9$^\circ$  & 30.3$^\circ$ \\
  \hline
  \end{tabular}
  \end{center}
  \label{table:skymap_stat}
\end{table*}

To compare BW's performance with LIB's and cWB's \citep{skyloc}, we created the equivalents of Figures 1-4 with LIB and cWB using the same subset of events we used in this study (see Figure Sets 1-4). We have found that all metrics show that these algorithms perform similarly in localizing the source. Histograms of $\mathcal{A}$ show that $\mathcal{A}$ values for BW are comparable to, but systematically bigger than for cWB and LIB for all morphologies, except for \textit{BBH} signals, for which BW typically yields smaller searched areas than LIB. Also, there are more \textit{WNB} skymaps with large searched areas ($\mathcal{A} \gtrsim 100$ deg$^2$) for LIB than for BW. This is likely due to its ability to recover more of the signal by using multiple wavelets as opposed to a single sine-Gaussian template.

\subsection{Waveform reconstruction} \label{ssec:wfrec}

BW uses sine-Gaussian wavelets to reconstruct a GW signal from the detector output, which means that the recovered signal is always given as a linear combination of sine-Gaussian wavelets, the number of which is a parameter in the RJMCMC. To characterize the quality of waveform reconstruction, we introduce the \emph{overlap} ($\mathcal{O}$, sometimes referred to as match) which measures the similarity of an injected ($h_{\rm i}$) and a recovered ($h$) waveform as:

\begin{equation}
\mathcal{O}=\frac{(h_{\rm i}|h)}{\sqrt{(h_{\rm i}|h_{\rm i}) (h|h)}}, 
\label{eq:overlap}
\end{equation}
where $(.|.)$ is a noise weighted inner product, defined as:

\begin{equation}
(a|b) = 2 \int_0^{\infty} \frac{a(f) b^*(f)+a^*(f) b(f)}{S_{\rm n} (f)} {\rm d} f,
\end{equation}
where $S_{\rm n}$ is the one-sided power spectral density of the detector noise, and $x^*$ denotes the complex conjugate of $x$.

From Eq. (\ref{eq:overlap}) it is visible that $\mathcal{O}$ ranges from -1 to 1, with $\mathcal{O}=1$ meaning perfect match between $h_i$ and $h$, $\mathcal{O}=0$ meaning no match at all, and $\mathcal{O}=-1$ meaning a perfect anti-correlation between $h_i$ and $h$. With Eq. (\ref{eq:overlap}), we can calculate the overlap using data from only one detector. To characterize the waveform reconstruction for the network of GW detectors, we introduce the \emph{network overlap} ($\mathcal{O}_{\rm net}$)  by changing the inner products in Eq. (\ref{eq:overlap}) with the sum of inner products calculated for different detectors:

\begin{equation}
\mathcal{O}_{\rm net} = \frac{\sum_{j=1}^N (h_{\rm i}^{(j)}|h^{(j)})}{\sqrt{\sum_{j=1}^N (h_{\rm i}^{(j)}|h_{\rm i}^{(j)}) \cdot \sum_{j=1}^N (h^{(j)}|h^{(j)})}},
\label{eq:netoverlap}
\end{equation}
where $j$ denotes the $j$-th detector in the network, and $N$ is the number of detectors used in the analysis (note that $N=2$ in this study). Note that in our analysis we only considered waveforms reconstructed from outputs of each detector ($h^{(j)}$), but not the astrophysical GW polarizations ($h_+$, $h_{\times}$), because the two polarizations cannot be decomposed from detections with two co-aligned GW detectors, such as H1 and L1.

\begin{figure}
\figurenum{5}
\epsscale{1.2}
\plotone{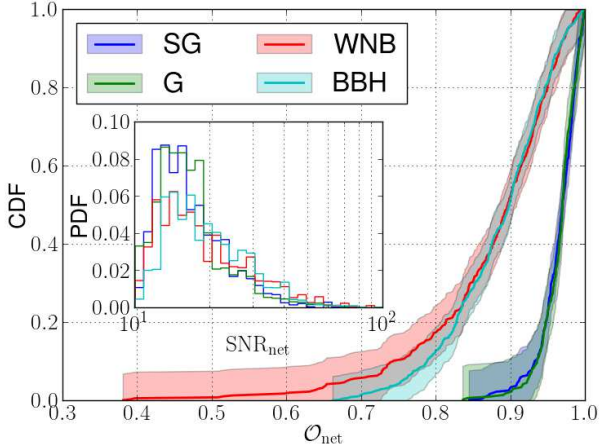}
\caption{Cumulative distribution function (CDF) of network overlaps ($\mathcal{O}_{\rm net}$). Shadings represent the 2$\sigma$ uncertainties calculated using the Dvoretzky--Kiefer--Wolfowitz inequality \citep{dkw-ineq}. The lower the curves reach at a given $\mathcal{O}_{\rm net}$ value, the better the reconstruction is. The inset shows the normalized histogram of network signal-to-noise ratio (SNR$_{\rm net}$) for signals with four different morphologies. The curves for \textit{SG} and \textit{G} signals are identical within the 2$\sigma$ statistical errors, and they indicate significantly better reconstructions of \textit{SG} and \textit{G} signals than of \textit{WNB} and \textit{BBH} signals. \label{fig:overlap_hist}}
\end{figure}

Figure \ref{fig:overlap_hist} shows the cumulative distribution functions (CDF) of $\mathcal{O}_{\rm net}$. Shaded ranges represent the 2$\sigma$ uncertainty calculated using the Dvoretzky--Kiefer--Wolfowitz inequality \citep{dkw-ineq}. The fraction of injected signals with $\mathcal{O}_{\rm net} > 0.9$ is 97\% for \textit{G}, 96\% for \textit{SG}, 48\% for \textit{BBH}, and 47\% for \textit{WNB} signals after the waveform reconstruction with BW. 95\% of injections have $\mathcal{O}_{\rm net} > 0.92$ for \textit{G} signals, $\mathcal{O}_{\rm net} > 0.91$ for \textit{SG} signals, $\mathcal{O}_{\rm net} > 0.75$ for \textit{BBH} signals, and $\mathcal{O}_{\rm net} > 0.68$ for \textit{WNB} signals. In Figure \ref{fig:overlap_hist}, the lower the curves reach at a given $\mathcal{O}_{\rm net}$ value, the better the reconstruction is. This suggests that BW's waveform reconstruction works most effectively for \textit{SG} and \textit{G} signals, for which the curves are identical within the 2$\sigma$ statistical error. BW's waveform reconstruction is less effective for \textit{WNB} and \textit{BBH} signals, and it shows similar characteristics for these morphologies at high network overlaps ($\gtrsim 0.8$), but the distribution for \textit{WNB} signals has a longer tail at low $\mathcal{O}_{\rm net}$ values. The better performance of BW for \textit{SG} and \textit{G} signals is due to the fact that at low SNR$_{\rm net}$ BW tends to use fewer wavelets to avoid overfitting the data. \textit{SG} and \textit{G} signals can be reconstructed accurately even with just 2-3 sine-Gaussian wavelets, while this is not possible for \textit{WNB} and \textit{BBH} signals. This also means that the curves for \textit{SG} and \textit{G} signals in Figure \ref{fig:overlap_hist} represent BW's maximal capability of reconstructing a GW signal for a given noise level, while the results for \textit{WNB} and \textit{BBH} signals represent BW's performance on more generic (and thus, more realistic) GW signals. Note that while $\mathcal{O}_{\rm net}$ values are smaller for \textit{WNB} and \textit{BBH} signals, BW detects these with more confidence, because its detection statistic has a stronger dependence on signal complexity than on SNR$_{\rm net}$ (for details see \citealt{BW_det}). The inset plot in Figure \ref{fig:overlap_hist} shows the normalized histogram of injected signals' network signal-to-noise ratio (SNR$_{\rm net}$) for the four different signal morphologies. \textit{SG} and \textit{G} signals have an overabundance at SNR$_{\rm net} \lesssim 20$ relative to \textit{WNB} and \textit{BBH} signals. This indicates that the previously described difference in the distribution of $\mathcal{O}_{\rm net}$ is not due to the different SNR$_{\rm net}$ distributions, as BW performs better for \textit{SG} and \textit{G} signals despite the fact that SNR$_{\rm net}$ values for \textit{SG} and \textit{G} signals are usually smaller than for \textit{WNB} and \textit{BBH} signals. Note that these distributions strongly depend on the parameter distributions of injected signals defined in Table 4 of \citet{skyloc}, and on the corresponding detection efficiencies of the combination of cWB and BW pipelines for the different parameter sets (see the SNR$_{\rm net}$ histogram in the inset of Figure \ref{fig:overlap_hist}).

\begin{figure*}[htbp!]
\figurenum{6}
\gridline{\fig{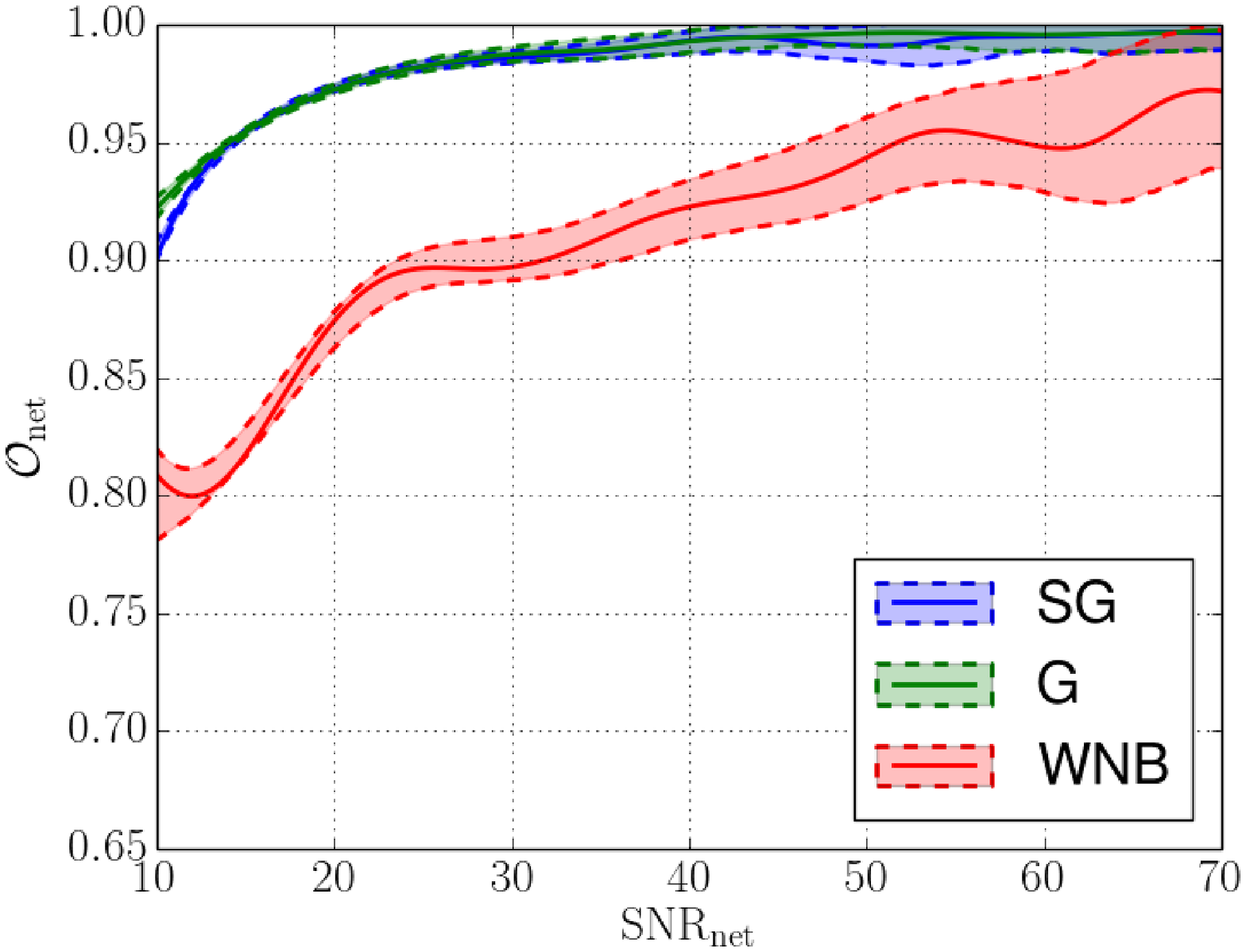}{0.5\textwidth}{}
		  \fig{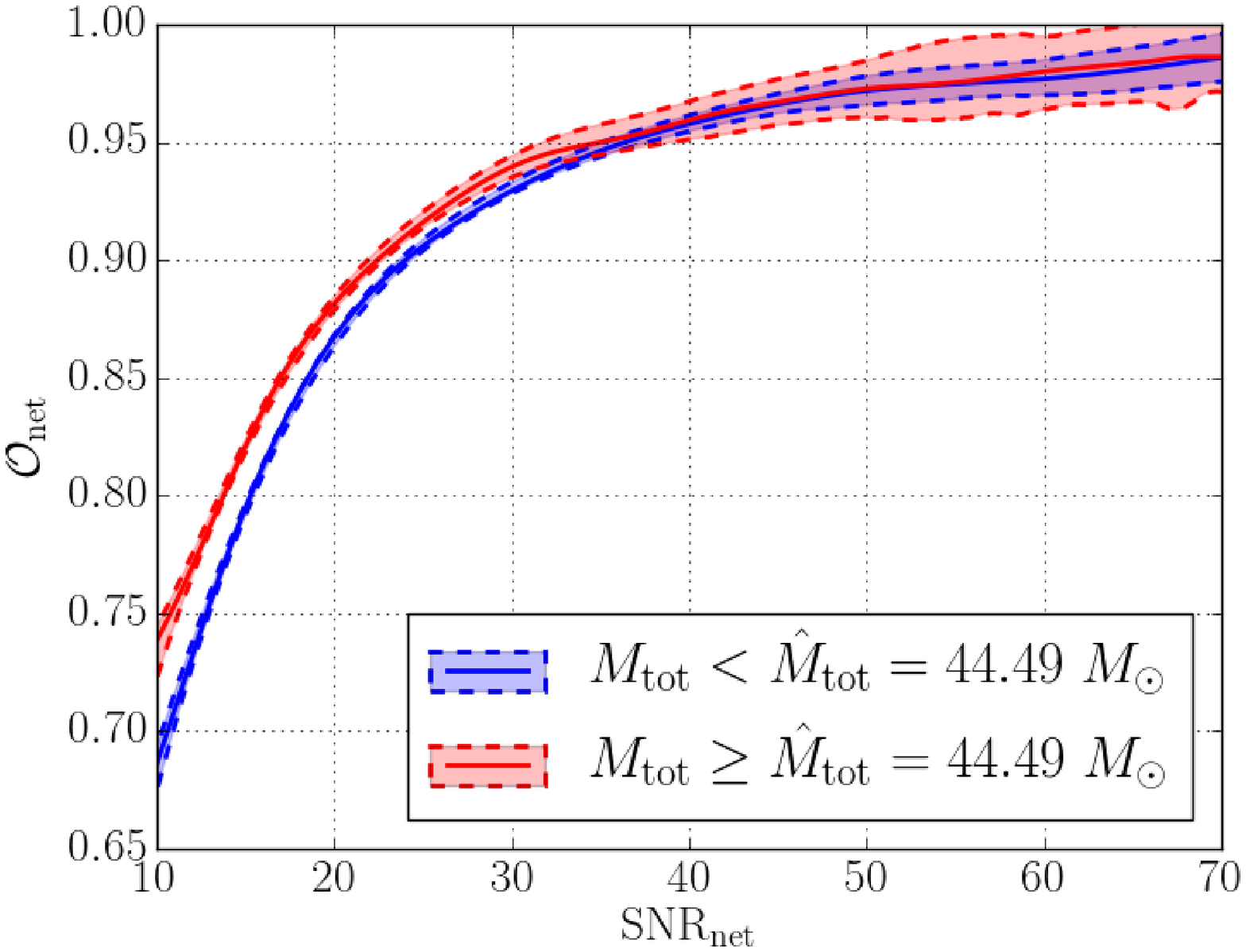}{0.5\textwidth}{}}
\caption{Dependence of network overlaps ($\mathcal{O}_{\rm net}$) on network signal-to-noise ratios (SNR$_{\rm net}$) for \textit{SG}, \textit{G}, \textit{WNB}, and \textit{BBH} signals. Note that we excluded the injections with SNR$_{\rm net}>$100 from the curve estimation. Shaded areas represent the 1$\sigma$ uncertainty regions of the measured $\mathcal{O}_{\rm net}$ values. Left panel shows SNR$_{\rm net}$ dependence of $\mathcal{O}_{\rm net}$ for \textit{SG}, \textit{G}, and \textit{WNB} signals. All three morphologies show a clear trend of increasing overlap with increasing SNR$_{\rm net}$. Right panel shows SNR$_{\rm net}$ dependence of network overlaps for \textit{BBH} signals with detector-frame total mass below and above the median total mass $\hat{M}_{\rm tot} = 44.49 \ M_{\odot}$. BW performed significantly better for signals with higher $M_{\rm tot}$ at SNR$_{\rm net}\lesssim 35$ values.\label{fig:overlap_snr}}
\end{figure*}

We show $\mathcal{O}_{\rm net}$ vs. SNR$_{\rm net}$ for \textit{SG}, \textit{G}, and \textit{WNB} signals in the left panel of Figure \ref{fig:overlap_snr}. Curves were estimated with a Gaussian kernel smoother, which is a nonparametric regression method. Shaded regions between dashed lines represent the 1$\sigma$ uncertainty regions calculated with the bootstrap method, in which we estimate the curve repeatedly for sub-samples randomly drawn from the full sample. Note that we excluded the injections with SNR$_{\rm net}>100$ from the estimation of these curves, and we only show the estimated curves up to SNR$_{\rm net}$=70. All three morphologies show a clear trend of $\mathcal{O}_{\rm net}$ increasing with SNR$_{\rm net}$.

For \textit{BBH} signals we calculated the $\mathcal{O}_{\rm net}$ vs. SNR$_{\rm net}$ curves in two separate bins of total mass ($M_{\rm tot}$) of the binary black hole system, calculated in the detector's frame. The two bins were defined with $M_{\rm tot}$ being $M_{\rm tot} < \hat{M}_{\rm tot}$ and $M_{\rm tot} > \hat{M}_{\rm tot}$, where $\hat{M}_{\rm tot} = 44.49 \ M_{\odot}$ is the median of $M_{\rm tot}$ values for all \textit{BBH} injections. The $\mathcal{O}_{\rm net}$ vs. SNR$_{\rm net}$ curves for \textit{BBH} signals are shown in the right panel of Figure \ref{fig:overlap_snr}. Similarly to other morphologies, \textit{BBH} injections also show a clear trend of increasing $\mathcal{O}_{\rm net}$ with increasing SNR$_{\rm net}$. At low ($\lesssim 35$) SNR$_{\rm net}$ values, BW performed significantly better for signals with higher $M_{\rm tot}$, while differences in the curves are within the level of statistical errors for higher SNR$_{\rm net}$ values. Signals with high $M_{\rm tot}$ are recovered with better accuracy because a large fraction of the signal power is in a compact region of time-frequency space and therefore can be captured with a small number of wavelets, while signals with low $M_{\rm tot}$ spend a comparatively longer amount of time in the sensitive band of the detectors, requiring more wavelets, and a larger total signal strength, to achieve a similar fit. This difference vanishes at high SNR$_{\rm net}$ because BW uses more wavelets to reconstruct signals with higher SNR$_{\rm net}$.

Similarly to Figure \ref{fig:overlap_hist}, Figure \ref{fig:overlap_snr} also shows that BW performs very similarly on \textit{SG} and \textit{G} signals, and much less efficiently on \textit{WNB} and \textit{BBH} signals. This is due to the fact that BW needs to use more wavelets to accurately reconstruct \textit{WNB} and \textit{BBH} signals. Note, that despite the weaker performance on \textit{WNB} and \textit{BBH} signals, these also approach the reconstruction accuracy for \textit{SG} and \textit{G} signals at higher SNR$_{\rm net}$ values. Comparing the two panels of Figure \ref{fig:overlap_snr}, it is visible that the curve for \textit{BBH} signals is similar to the curve for \textit{WNB} signals, with slightly worse overlap at low SNR$_{\rm net}$ and slightly better overlap at high SNR$_{\rm net}$ values.

Our results show that BW robustly reconstructs waveforms with various morphologies. Although there are significant differences between the efficiency of reconstructions of signals with different morphologies, even for the worst case of \textit{WNB} signals (which do not even match BW's assumption of the signal always being elliptically polarized), most of them have relatively high overlaps, and there is a clear trend of $\mathcal{O}_{\rm net}$ approaching 1 as SNR$_{\rm net}$ increases.

\subsection{Point estimates of waveform central moments} \label{ssec:pointest}

For a generic burst signal, we do not have any specific astrophysical model whose parameters could be estimated. In this case we can still give estimates on model-independent parameters of the signal. Here we consider the central moments of the waveform as such parameters.

The first central moments are central time ($t_0$) and central frequency ($f_0$), and the second central moments are duration ($\Delta t$) and bandwidth ($\Delta f$), defined as

\begin{subequations}
\begin{equation}
t_0 = \int_{-\infty}^{\infty} {\rm d}t \ \rho _{\rm TD}(t) t
\label{eq:t0}
\end{equation}
\begin{equation}
f_0 = \int_{0}^{\infty} {\rm d}f \ \rho_{\rm FD}(f) f
\label{eq:f0}
\end{equation}
\begin{equation}
(\Delta t)^2 = \int_{-\infty}^{\infty} {\rm d}t \ \rho_{\rm TD}(t) (t - t_0)^2
\label{eq:dt}
\end{equation}
\begin{equation}
(\Delta f)^2 = \int_{0}^{\infty} {\rm d}f \ \rho_{\rm FD}(f) (f - f_0)^2
\label{eq:df}
\end{equation}
\end{subequations}
respectively, where $\rho_{\rm TD}$ and $\rho_{\rm FD}$ are the effective normalized distributions of signal energy, expressed in the time domain (TD) and in the frequency domain (FD):

\begin{subequations}
\begin{equation}
\rho_{\rm TD}(t) = \frac{h(t)^2}{h_{\rm rss}^2},
\end{equation}
\begin{equation}
\rho_{\rm FD} (f) = \frac{2 (|\tilde{h}(f)|^2}{h_{\rm rss}^2},
\end{equation}
\end{subequations}
where $h(t)$ is the  whitened (i.e.~normalized with the amplitude spectral density of the detector noise) waveform for a given detector and $\tilde{h}(f)$ is the Fourier transform of $h(t)$. These distributions satisfy $\int_{-\infty}^{\infty} \rho_{\rm TD}(t) {\rm d}t = 1$, and $\int_{0}^{\infty} \rho_{\rm FD}(f) {\rm d}f = 1$.

Estimations of higher order moments could also be given with BW, however we excluded them from our analysis due to the fact that they are more strongly affected by statistical errors than estimations of the first order moments (for a detailed discussion of this, see the end of this section).

BW reconstructs the waveform and calculates the waveform moments for each sample in the Markov chain. We calculated the median value to give a point estimate of the waveform moments. To quantify the accuracy of the point estimate of waveform moment $x$, we define the absolute error of the estimation, $e_x$, as:

\begin{equation}
e_x = |x^{\rm (e)} - x^{\rm (r)}|,
\end{equation}
where $x^{\rm (e)}$ is the estimated, and $x^{\rm (r)}$ is the real value of $x$. We also introduce the relative error of an estimate, $\eta_x$, as:

\begin{equation}
\eta_x = \frac{e_x}{x^{\rm (r)}}.
\end{equation}

\begin{figure*}[htbp]
\figurenum{7}
\gridline{\fig{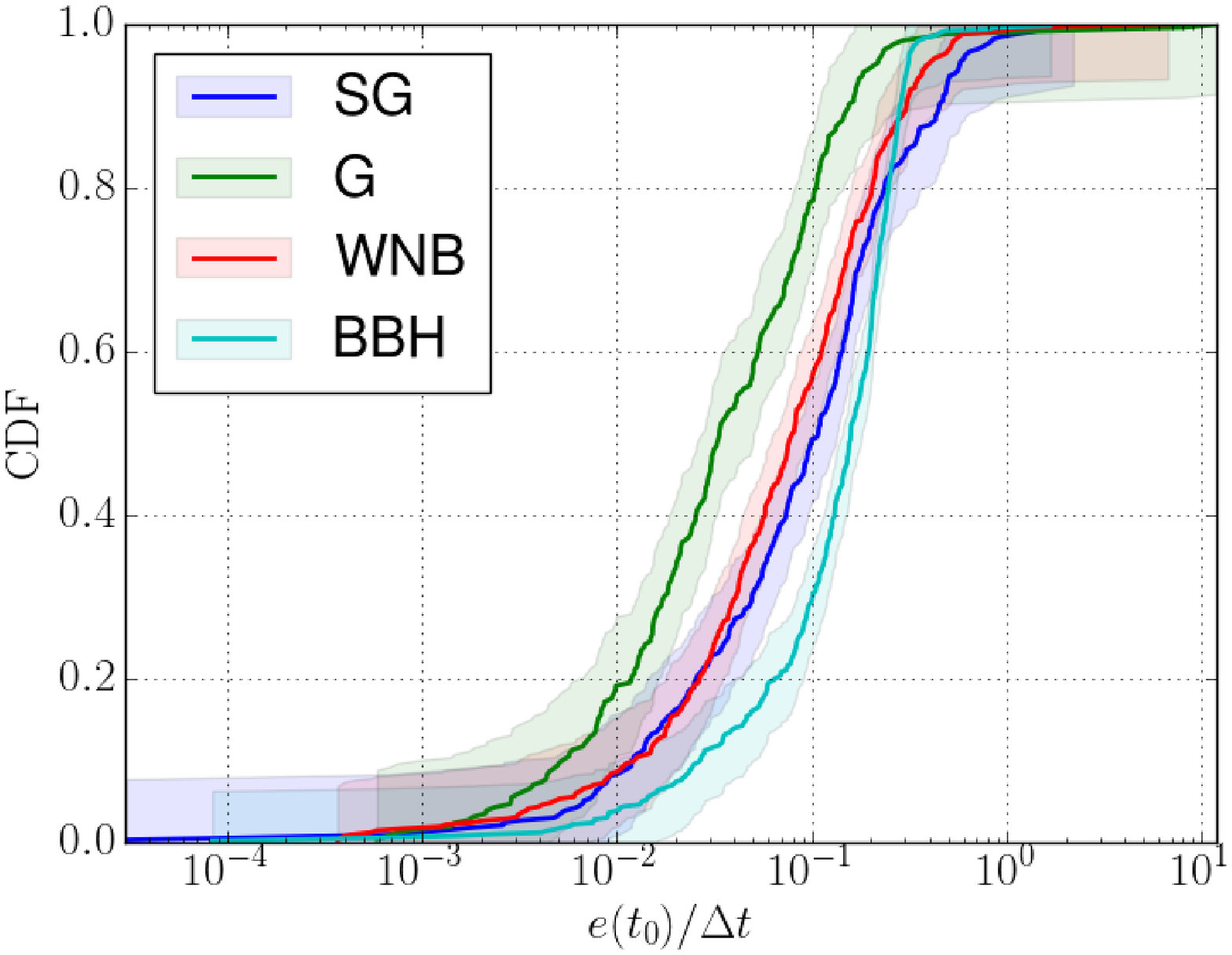}{0.5\textwidth}{}
		  \fig{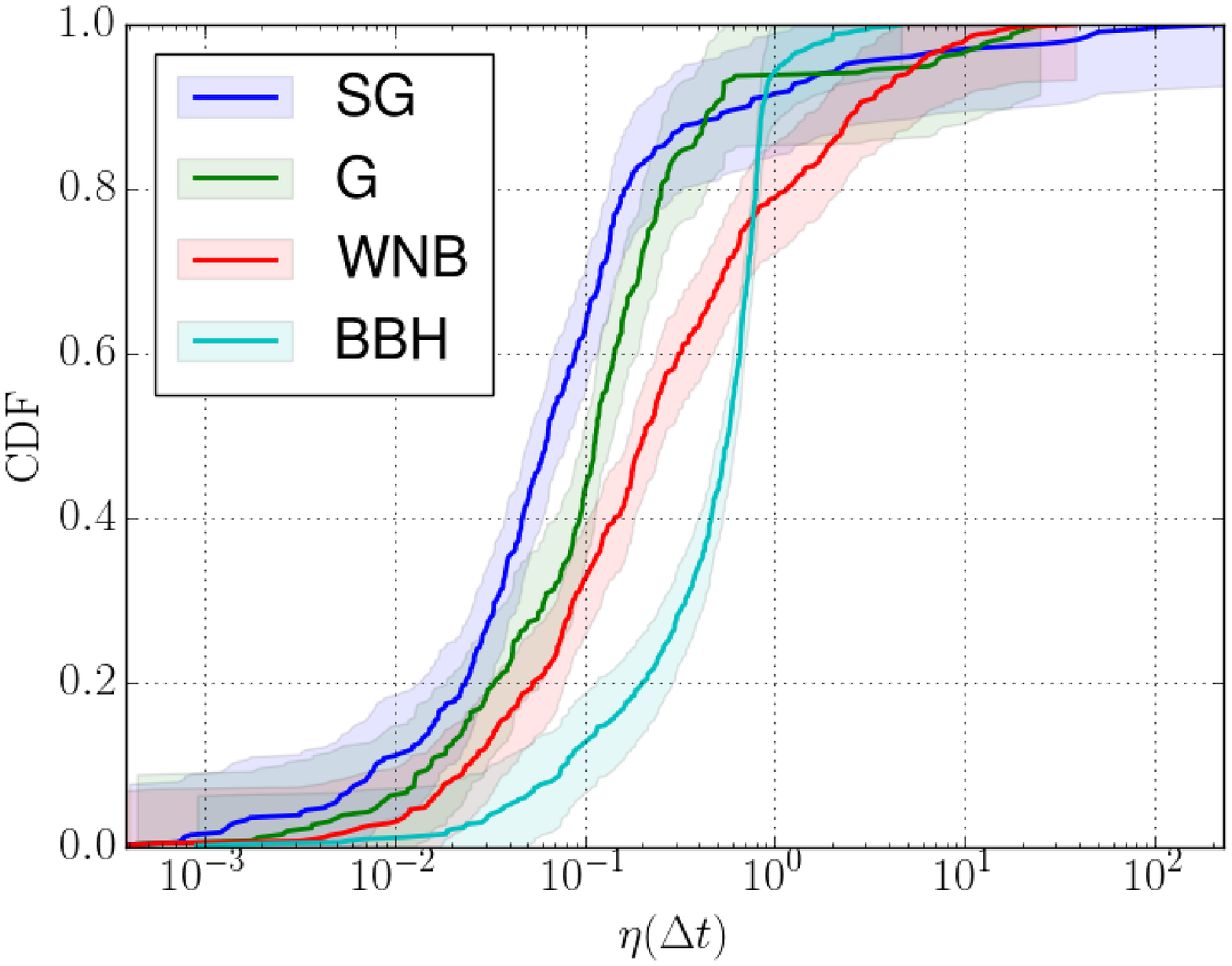}{0.5\textwidth}{}}
\gridline{\fig{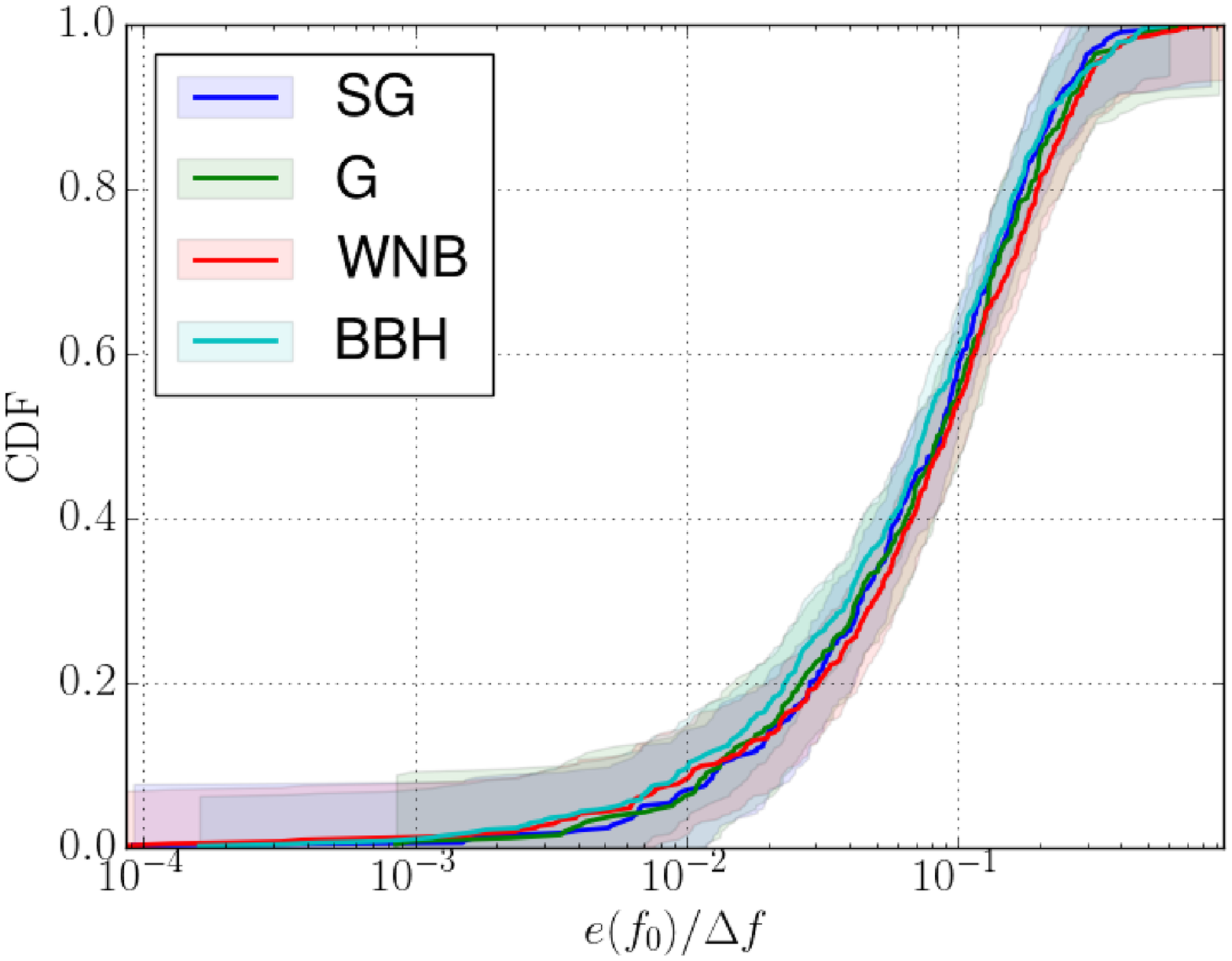}{0.5\textwidth}{}
          \fig{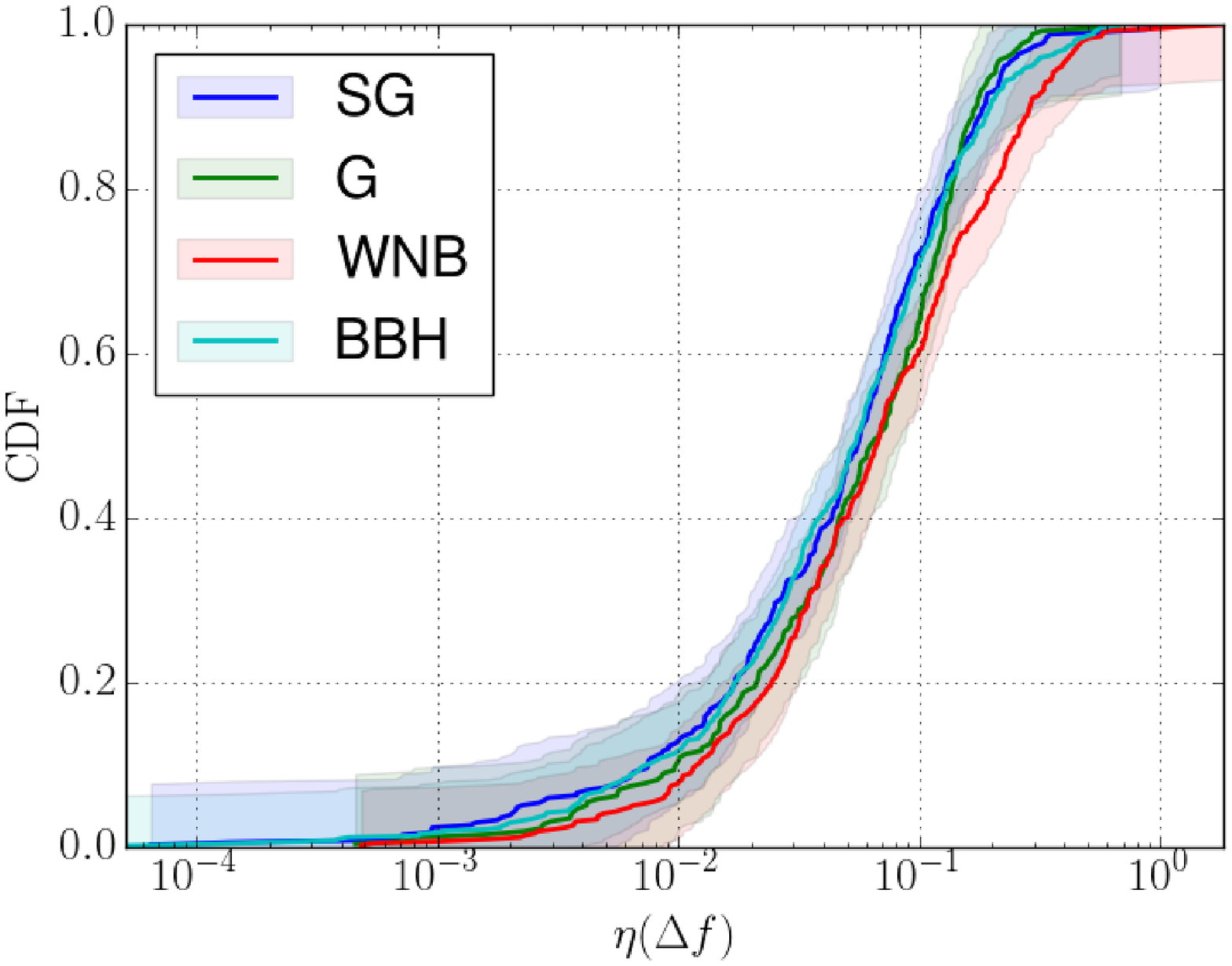}{0.5\textwidth}{}}
\caption{Cumulative distribution functions (CDF) of waveform central moment errors: absolute errors of central time estimations divided by signal durations ($e_{t_0}/\Delta t$, upper left), relative errors of duration estimations ($\eta_{\Delta t}$, upper right), absolute errors of central frequency estimations divided by signal bandwidths ($\eta_{f_0}/\Delta f$, lower left), and relative errors of bandwidth estimations ($\eta_{\Delta f}$, lower right). Shadings represent the 2$\sigma$ uncertainties calculated using the Dvoretzky--Kiefer--Wolfowitz inequality \citep{dkw-ineq}. Colors indicate CDFs for signals with sine-Gaussian (\textit{SG}), Gaussian (\textit{G}), white noise burst (\textit{WNB}), and binary black hole (\textit{BBH}) morphologies. We give values of 95th percentiles and medians in Table \ref{tab:errors}.  \label{fig:estim_err}}
\end{figure*}

We show CDFs of $e_{t_0}/\Delta t$, $e_{f_0}/\Delta f$, $\eta_{\Delta t}$, and $\eta_{\Delta f}$ in Figure \ref{fig:estim_err}, where shadings represent the 2$\sigma$ uncertainties calculated using the Dvoretzky--Kiefer--Wolfowitz inequality \citep{dkw-ineq}. All moments were calculated for H1 detector data, however, results are very similar for L1 too. We divided the absolute errors of the first moment estimations with the real values of the corresponding second moments, because we expect that the statistical error of first moment estimation scales with the real values of the second moments.

We show CDFs of $e_{t_0}/\Delta t$ for different morphologies in the top left panel of Figure \ref{fig:estim_err}. These show that the most accurate $t_0$ estimates with BW are obtained for \textit{G} signals, while estimates for \textit{BBH} signals are the least accurate. The relatively large $e_{t_0}$ values are due to the fact that BW cannot reconstruct low-amplitude parts of the signal overwhelmed by noise, which can cause a systematic error in the estimation of $t_0$. For example, BW is almost insensitive to the inspiral parts of \textit{BBH} signals, which make up the bulk of \textit{BBH} signal durations, and this bias increases the smaller the total mass of the systems are. This effect is less significant for the other three morphologies, which explains why the estimation of $t_0$ is less accurate for \textit{BBH} signals (with a median $e_{t_0}$ value of $0.16 \Delta t$). The $t_0$ values we obtain for H1 and for L1 are strongly correlated, which means that the error on the estimation of the difference of arrival times between H1 and L1 (determining the thickness of the sky localization triangulation ring) is typically smaller than $e_{t_0}$.

We show CDFs of $\eta_{\Delta t}$ in the top right panel of Figure \ref{fig:estim_err}. These curves significantly differ for different waveform morphologies. Regarding the median $\eta_{\Delta t}$, $\Delta t$ estimation is the most accurate for \textit{SG} signals (with a median value of 0.06) and the least accurate for \textit{BBH} signals (with a median value of 0.57). Note however, that median values do not tell about how  long the tails of $\eta_{\Delta t}$ distributions are. From the four morphologies, the $\eta_{\Delta t}$ distribution for the \textit{SG} signals have the longest tail (see top right panel of Figure \ref{fig:estim_err}). For $\eta_{\Delta t} \lesssim 1$, CDF values for \textit{BBH} signals are significantly smaller than for the other three morphologies, while for $\eta_{\Delta t} \gtrsim 1$, they are higher. This is due to the steep part of the \textit{BBH} curve around $\eta_{\Delta t} = 1$, which corresponds to the systematic underestimation of the duration of low mass \textit{BBH} signals (due to the effect explained in the previous paragraph).

We show CDFs of $e_{f_0}/\Delta f$ in the bottom left panel of Figure \ref{fig:estim_err}. Curves for different morphologies are identical within the error bars, in contrast with CDFs of $e_{t_0}/\Delta t$ where the curves are similar but not identical. This indicates that these errors are purely due to the statistical errors of central frequency estimation, determined by the non-zero value of $\Delta f$. Note that all $e_{f_0}$ values are smaller than $\Delta f$, and the median of $e_{f_0}$ is smaller than 0.1 for all morphologies (see Table \ref{tab:errors}).

We show CDFs of $\eta_{\Delta f}$ in the bottom right panel of Figure \ref{fig:estim_err}. The accuracies of $\Delta f$ estimation are similar for different morphologies, but not as much as for $\eta_{f_0}/\Delta f$. 95th percentiles are between 0.2 and 0.4 for the different morphologies. Note that relative errors of bandwidth estimations tend to be higher than of central frequency estimations. This is due to the fact that estimations of second order moments inherit errors from estimations of lower order moments (see Eq. (\ref{eq:dt}) and (\ref{eq:df})), and thus have higher statistical errors. We expect that estimation of third and higher order moments would have even bigger errors, and thus we restrict our attention to examining only estimations of the first two moments. Medians and 95th percentiles of errors for each moment and for each morphology are shown in Table \ref{tab:errors}.

\floattable
\begin{deluxetable}{c|ccccc}
\tablecaption{Medians (50th percentiles) and 95th percentiles of waveform central moment errors for the \textit{SG}, \textit{G}, \textit{WNB}, and \textit{BBH} signal morphologies. $P$ denotes the percentile rank of values given in the corresponding table columns.\label{tab:errors}}
\tablecolumns{4}
\tablewidth{0pt}
\tablehead{
\colhead{ } & \colhead{$P$} &  \multicolumn4c{Signal morphology} \\
\cline{3-6}
\colhead{ } & \colhead{ } & \colhead{\textit{SG}} & \colhead{\textit{G}} & \colhead{\textit{WNB}} & \colhead{\textit{BBH}} 
}
\startdata
$e_{t_0}/\Delta t$ & 50th & 0.11 & 0.03 & 0.08 & 0.16 \\
{    }   & 95th & 0.57 & 0.21 & 0.39 & 0.31 \\
\hline
$\eta_{\Delta t}$ & 50th & 0.06 & 0.11 & 0.21 & 0.57 \\
{    }           & 95th & 2.30 & 6.59 & 5.60 & 1.07 \\
\hline
$e_{f_0}/\Delta f$ & 50th & 0.09 & 0.09 & 0.09 & 0.07 \\
{    }      & 95th & 0.29 & 0.30 & 0.32 & 0.31 \\
\hline
$\eta_{\Delta f}$ & 50th & 0.06 & 0.07 & 0.07 & 0.06 \\
{    }           & 95th & 0.23 & 0.21 & 0.39 & 0.30 \\
\enddata
\end{deluxetable}

As a summary, results presented in Figure \ref{fig:estim_err} show that the distributions of errors for $f_0$ and $\Delta f$ are very similar for different morphologies, while distributions of errors for $t_0$ and $\Delta t$ show significant differences between different morphologies. This also means that while errors of $f_0$ and $\Delta f$ estimations are purely statistical, errors of $t_0$ and $\Delta t$ estimations include systematics as well. The latter is due to the fact that BW cannot reconstruct low-amplitude parts of a signal overwhelmed by noise, which may result with a systematic error in the estimation of $t_0$ and $\Delta t$. It is clear that the accuracy of moment estimation is affected by how accurately signals are reconstructed. However, we see identical CDFs of $e_{f_0}/\Delta f$ for different morphologies, while these have different $\mathcal{O}_{\rm net}$ distributions, which suggests that $\mathcal{O}_{\rm net}$ is not a good indicator of BW's moment estimation accuracy.

\section{Conclusion} \label{sec:concl}
We presented a comprehensive multi-aspect study on the performance of BW, a Bayesian GW burst PE pipeline used by the LIGO-Virgo Collaboration for reconstructing GW burst signals and their parameters. We injected a large number of simulated signals with four different morphologies (sine-Gaussians, Gaussians, white-noise bursts, and binary black hole signals) into simulated O1 aLIGO noise to test BW's performance in three different aspects of PE: sky localization, waveform reconstruction, and estimation of waveform central moments (for details on the methods we used see Section \ref{sec:methods}).

BW localizes sources with a level of accuracy comparable for all four morphologies, with the median separation of actual and estimated sky locations ranging from 25.1$^{\circ}$ to 30.3$^{\circ}$ (see Table \ref{table:skymap_stat}), and median searched area ($\mathcal{A}$, see Eq. \ref{eq:searchedarea}) ranging from 99.2 deg$^2$ to 252.8 deg$^2$ (see Section \ref{ssec:skyloc}). This is reasonable accuracy for a two-detector network, and is comparable to accuracies of other localization pipelines (cWB and LIB) studied previously \citep{skyloc}. Histograms of $\mathcal{A}$ (see Figure \ref{fig:searched_area}) show that $\mathcal{A}$ values for BW are comparable to, but systematically bigger than for cWB and LIB for all morphologies. The exceptions are \textit{BBH} signals, for which BW's $\mathcal{A}$ values are systematically smaller. Note that the runtime of cWB and LIB is much shorter than of BW.

BW reconstructs waveforms as a linear combination of sine-Gaussian wavelets. To measure the goodness of reconstruction, we used the network overlap ($\mathcal{O}_{\rm net}$, see Eq. \ref{eq:netoverlap}), which quantifies the similarity between the injected and the reconstructed signals. We have found that BW reconstructs signals with $\mathcal{O}_{\rm net}>0.9$ for 98\% of \textit{G}, 96\% of \textit{SG}, 45\% of \textit{WNB}, and 47\% of \textit{BBH} signals (see Section \ref{ssec:wfrec}). We have also found that (see Figure \ref{fig:overlap_snr}) $\mathcal{O}_{\rm net}$ increases rapidly with increasing SNR$_{\rm net}$, reaching $\mathcal{O}_{\rm net}=0.95$ at SNR$_{\rm net} \approx 14$ for \textit{SG} and \textit{G}, at SNR$_{\rm net} \approx 50$ for \textit{WNB}, and at SNR$_{\rm net} \approx 35$ for \textit{BBH} signals. These results suggest that we can expect very good reconstruction ($\mathcal{O}_{\rm net}>0.95$) for almost any signal with high ($\gtrsim 50$) SNR$_{\rm net}$, and reasonably good reconstruction ($\mathcal{O}_{\rm net}>0.85$) for almost any signal with moderate ($\gtrsim 20$) SNR$_{\rm net}$. 

We also examined how accurately BW can estimate the central moments of a GW waveform (see Section \ref{ssec:pointest}). These are model-independent parameters of a signal, therefore by examining the estimation of them, we can characterize PE without assuming any astrophysical model for the source. We have found that errors of $f_0$ and $\Delta f$ estimations are purely statistical, while errors of $t_0$ and $\Delta t$ estimations include some systematics as well. We have also found that $\mathcal{O}_{\rm net}$ is not a good indicator of BW's moment estimation accuracy. The median value of $e_{f_0}/\Delta f$ is 0.09 for \textit{SG}, \textit{G} and \textit{WNB} signals, and 0.07 for \textit{BBH} signals (see Table \ref{tab:errors}). There is no standard procedure on how the estimated moments of GW bursts can be used to test astrophysical models, however future studies can use our results to test the feasibility of particular methods using signal moments.

This paper fits into a series of studies examining PE for GW bursts (see e.g.~\citealt{cWB_PE, skyloc}). These studies can be used in comparisons with improved performances of future PE pipelines, and in testing the feasibility of possible astrophysical applications of future GW burst detections.

\acknowledgments
This paper was reviewed by the LIGO Scientific Collaboration under LIGO Document P1600181. We thank Marco Drago and Sergey Klimenko for their valuable comments on the manuscript. We acknowledge the Burst First2Years sky localization Open Data release\footnote{\url{http://www.ligo.org/scientists/burst-first2years/}}. The authors acknowledge the support of the National Science Foundation and the LIGO Laboratory. LIGO was constructed by the California Institute of Technology and Massachusetts Institute of Technology with funding from the National Science Foundation and operates under cooperative agreement PHY-0757058. The authors would like to acknowledge the use of the LIGO Data Grid computer clusters for performing all the computation reported in the paper. Bence B\'ecsy was supported by the \'UNKP-16-2 New National Excellence Program of the Ministry of Human Capacities. Bence B\'ecsy was supported by the Hungarian Templeton Program that was made possible through the support of a grant from Templeton World Charity Foundation, Inc. The opinions expressed in this publication are those of the authors and do not necessarily reflect the views of Templeton World Charity Foundation, Inc. Peter Raffai is grateful for the support of the Hungarian Academy of Sciences through the "Bolyai J\'anos" Research Scholarship programme.

\appendix

\section{Resolving BayesWave O1 version's issue with high-q signals} \label{sec:app}
During O1, BW was prone to classifying simulated short-duration high-frequency signals which underwent many wave cycles (i.e. high-$Q$ signals) while in the measurement band of the detector as glitches.  In principle there is no reason for the Bayesian evidence used to rank hypothesis under consideration by BW to have strong frequency dependence. 

Upon examination of the mis-classified injections, it was revealed that the high-$f$, high-$Q$ signals exhibit multimodal likelihood support in the ($\alpha, \delta, t_0, f_0$) parameter sub-space. For these signals, the Markov Chain Monte Carlo (MCMC) sampler, which serves as the central engine to the BW algorithm, was not generically sampling between the different modes, and was thus prone to missing significant portions of the coherent signal and preferring the incoherent glitch model (which does not suffer the correlations between time-frequency parameters and sky location).

The cause of the multimodal likelihood function is clear.  For a sinusoidal signal ($Q=\infty$) the waveform is perfectly degenerate when time-shifted by an integer number of wave-periods ($T$). For high-$Q$ signals, a number of integer periods (or half-integer periods with a $\pi$ radians phase shift), time shifts produce similarly good fits to the data.  For coherent signals, these (nearly) degenerate time shifts are also present in the time delay between detectors, which, for BW, is encoded in the sky location.

To overcome BW's susceptibility to missing modes of the likelihood when analyzing high-$Q$ signals, we added a proposal distribution to the MCMC which explicitly suggests half-integer-period time shifts, along with half-integer-cycle phase shifts, for the wavelet parameters.  Furthermore, extensive development (beyond the scope of this paper) to improve the overall capabilities of BW's MCMC to sample the complicated sky-location posteriors encountered by two-detector gravitational-wave networks has been completed.

\begin{figure}[htbp!]
\figurenum{8}
\includegraphics[angle=270, width=\textwidth]{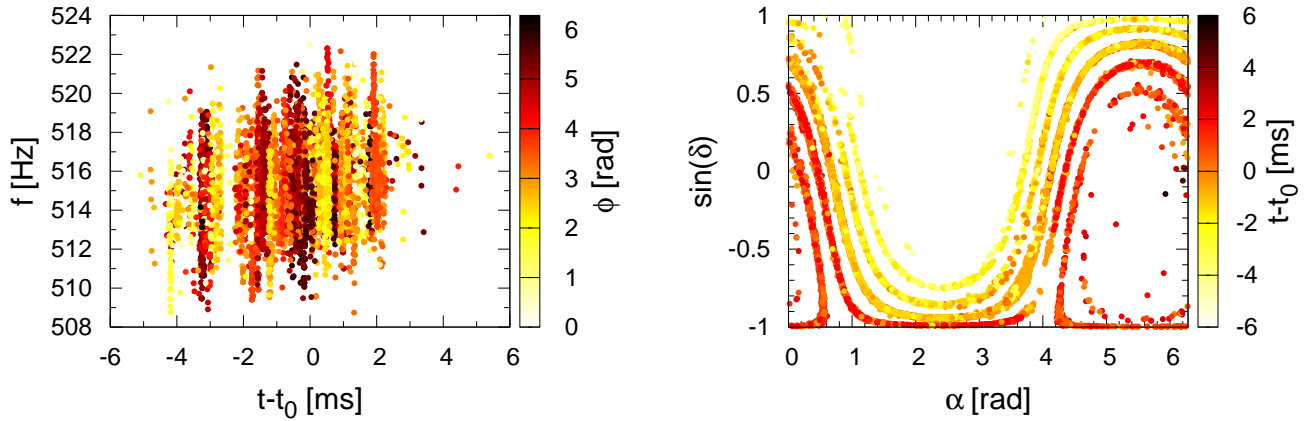}
\caption{Scatter plot of MCMC samples for signal model parameters of a high-$Q$, high-$f$ sine-Gaussian injection.  The left panel shows the time-frequency plane with points colored by the wavelet phase parameter.  Multiple modes and their phase-dependence are evident.  The right panel shows the same chain samples, but now projected on the the sky-location plane of the parameter space and colored by the time parameter.  Here again it is plain to see how different half-integer-period time shifts correspond to different ``rings'' on the sky, making this a challenging distribution to sample without well-tuned proposal distributions. \label{fig:frequency-phase-proposal}}
\end{figure}

Figure~\ref{fig:frequency-phase-proposal} contains two scatter plots from the BW MCMC utilizing the dedicated proposal distributions.  The multi-modal nature of the posterior is on clear display, as is the efficiency with which the MCMC sampler is able to move between local maxima in the likelihood.  This example came from an $f\sim512$ Hz, $Q\sim40$ sine-Gaussian injection.  Using the MCMC as it was during O1 we found a preference for the incoherent ``glitch'' model, with a Bayes factor between that and the coherent ``signal'' model of $\sim e^{60}$ in favor of the glitch model.  Using the updated sampler and analyzing the same (simulated) data we find a Bayes factor of $\sim e^{18}$ in favor of the signal model.

Despite this upgrade to BW's MCMC engine, we elected to present results \emph{as the algorithm performed} during O1 to facilitate a direct comparison with the snapshot of other burst parameter estimation techniques during the first observing run.  Future studies showing how the upgraded sampler performs on similar injections are underway.

\section{Example skymaps} \label{sec:exsky}

\begin{figure}
\figurenum{9}
\plotone{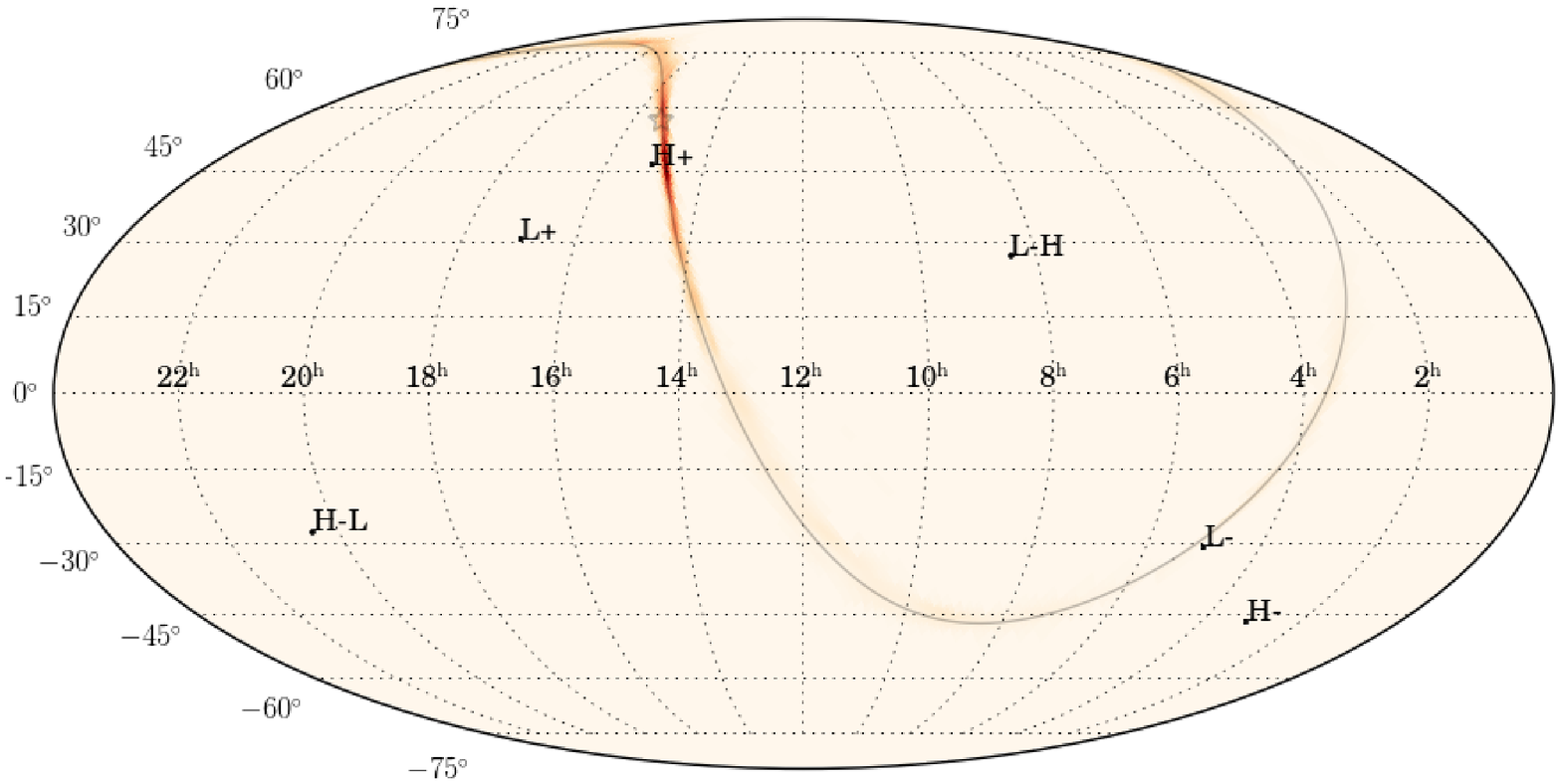}
\caption{An example skymap showing the reconstructed sky location for an injected \textit{SG} signal. The injected location is marked with a star and the corresponding triangulation ring for L1 and H1 detectors is denoted with a grey line. H-L and L-H marks the direction between the two detectors, H+ and L+ the directions above the detectors, and H- and L- the directions below the detectors. The complete figure set (20 figures) showing 5 example skymaps for each morphology is available in the online journal. Skymaps for all the signals used in this study are available in the Burst First2Years sky localization Open Data release (\url{http://www.ligo.org/scientists/burst-first2years/}). \label{fig:skymap}}
\end{figure}

Figure \ref{fig:skymap} shows an example skymap for an injected \textit{SG} signal. The injected location is marked with a star and the corresponding triangulation ring for L1 and H1 detectors is denoted with a grey line. H-L and L-H marks the direction between the two detectors, H+ and L+ the directions above the detectors, and H- and L- the directions below the detectors. Skymap in Figure \ref{fig:skymap} is a typical one. It is consistent with the triangulation ring of the two detector network and the constraint of the network antenna pattern, which leads to a relatively small elongated area on the sky with the maximum close to the injected location. Figure Set 8 shows 20 example skymaps (5 for each morphology) in the online journal. Skymaps for all the signals used in this study are available in the Burst First2Years sky localization Open Data release\footnote{\url{http://www.ligo.org/scientists/burst-first2years/}}.

\bibliography{mybib2}{}
\bibliographystyle{aasjournal}

\end{document}